\documentclass[twocolumn,showpacs,preprintnumbers,amsmath,amssymb,epsfig]{revtex4}

\usepackage{graphicx}
\usepackage{dcolumn}
\usepackage{bm}
\usepackage{epsfig}

\def\ga{\mathrel{\mathpalette\fun >}}
\def\fun#1#2{\lower3.6pt\vbox{\baselineskip0pt\lineskip.9pt
  \ialign{$\mathsurround=0pt#1\hfil##\hfil$\crcr#2\crcr\sim\crcr}}}

\input epsf

\def\plotone#1{\centering \leavevmode
\epsfxsize= 1.0\columnwidth \epsfbox{#1}}

\newcommand{\tableskip}{\\[-6pt]}
\def\be{\begin{equation}}
\def\ee{\end{equation}}
\def\ba{\begin{eqnarray}}
\def\ea{\end{eqnarray}}

\def\nn{\nonumber}

\begin{document}

\preprint{}

\title{Looking for an extra dimension with tomographic cosmic shear}

\author{Yong-Seon Song}
\email{ysong@cfcp.uchicago.edu}
\affiliation{
Enrico Fermi Institute, University of Chicago, Chicago IL 60637 
}

\date{\today}

\begin{abstract}
The cosmic acceleration was discovered in one of the brane-based models.
We are interested in discriminating this model from 
the dark energy by tomographic cosmic shear.
Growth factors are different in the two models  
when one adjusts parameters to get nearly identical $H(z)$.
The two models could be distinguished with independent determinations of
both geometrical factors and the growth factor.
We introduce new parameterizations to separate 
the influence of geometry and the influence of growth on cosmic shear maps.
We find that
future observations will be able to distinguish between both models.
\end{abstract}

\pacs{draft}


\maketitle

\section{Introduction}
The discovery of cosmic acceleration \cite{riess98,perlmutter98,perlmutter99} 
has led to 
theoretical efforts to understand the nature of the cosmic acceleration
\cite{ratra88,wang99,dvali00a,carroll00}
and phenomenological efforts to discriminate the diverse modelings
\cite{weller00,linder02,huterer02,weller03,bean03,song03b,kratochvil04}.
Most theoretical explanations for the acceleration preserve Einstein 
gravity and add a new smooth and dark component called Dark Energy (DE)
\cite{ratra88,wang99}.
However, DE is not yet strongly supported
theoretically nor phenomenologically.
The similar cosmic acceleration was also discovered 
in one of brane-based models \cite{dvali00a}
by C.~Deffayet et.al. \cite{deffayet00a,deffayet01a} 
(hereafter DGP model represents this discovery of the 
cosmic acceleration in the brane-based model \cite{dvali00a}.) 

In the DGP model,
the cosmic acceleration is generated by the modified gravity 
due to the presence of the extra dimension in DGP. 
The study of the extra dimension physics influencing the DGP model
has been investigated
by \cite{arkani-hamed98,randall99a,randall99b}.
Dvali, Gabadadze and Porrati designed a brane where
ordinary matter is embedded in an infinite 
extra dimension and gravity is modified at larger scale \cite{dvali00a}.
The model (DGP) which generates the cosmic acceleration based upon 
\cite{dvali00a}-type brane model was proposed in
\cite{deffayet00a,deffayet01a}.

While the DE growth factor is suppressed by dark energy component
domination over matter component, the DGP growth factor is suppressed
by the weakened gravity.
We derive the DGP growth factor by using the set of continuity
equations of ordinary matter confined to the DGP brane.
We find that the DGP growth factor departs noticeably from the DE growth factor
even with nearly identical expansion rate $H(z)$.
A slight difference is noticed from previous work
based upon a different idea \cite{lue04}.

The constraints from cosmic shear maps on DE 
with a model dependent parametrization
have been studied in 
\cite{hu99b,hu02a,jain03,bernstein03,takada03,song03b,zhang03}.
The cosmological parameters related to DE are determined by
a distinct evolution of
a geometrical factor and a growth factor. 
We introduce a more general parametrization to determine directly,
from cosmic shear data,
the distance and the growth factor as a function of redshift.
Both quantities can be precisely determined by the combination of 
the future CMB experiments and the future cosmic shear
surveys.
We investigate how to probe the difference between
the DGP growth factor and the DE growth factor 
with $H(z)$ determined through its geometrical influence on shear data.

\section{DGP model}
We briefly review the DGP model and set up a formalism to draw our result.
We consider a (4+1)-dimensional model in which 
no energy sources are present at an infinite bulk dimension
except at a (3+1)-dimensional brane embedded in the bulk.
The Einstein-Hilbert action is 
\ba\label{5Daction}
S_{(5)}=-\frac{M_{(5)}^3}{16\pi}\int d^5x\sqrt{-g_{(5)}}R_{(5)},
\ea
where the subscript $(5)$ denotes 
that quantities are five-dimensional, and
$M_{(5)}$ is the Planck mass in (4+1)-dimension.
We add a (3+1)-dimensional brane with matter fields and
the induced metric to the action Eq.~\ref{5Daction}.
The embedded three-brane action is \cite{dvali00a}
\ba\label{4Daction}
S_{(4)}=\int d^4x\sqrt{-g_{(4)}}
\left({\cal L}_m-\frac{M_{(4)}^2}{16\pi}R_{(4)}\right),
\ea
where the subscript $(4)$ denotes 
that quantities are four-dimensional, and
$M_{(4)}$ is the Planck mass in (3+1)-dimension.

We introduce a five dimensional metric with a line element
\ba
ds_{(5)}^2=-n^2(t,y)dt^2+a^2(t,y)\delta_{ij}dx^idx^j+dy^2,
\ea
where a flat metric is used for spatial dimensions and
the coordinate $y$ represents the extra dimension. 
Our universe is located at hypersurface $y=0$
where a lapse function is given by $n(t,0)=1$ and 
a spatial expansion factor $a(t,0)$ is determined by
the cosmic expansion of our universe 
\cite{deffayet00a}.

In the DGP model, the gravitational force at short distances
smaller than the crossover length $r_c$ scales as $1/r^2$
and the gravitational force at large distances
bigger than $r_c$ scales as $1/r^3$.
The crossover length scale $r_c$ is given by
\ba
r_c=\frac{M_{(4)}^2}{2M_{(5)}^3}.
\ea
When the Hubble horizon $H^{-1}$ is close to $r_c$,
the gravity is weakened and the cosmic expansion is accelerated.
It is an alternative explanation of the cosmic acceleration
with the absence of a dark energy component.
We investigate the phenomenological consequences
of a DGP-type extra dimension
in the following subsections, such as the geometrical factor
and the growth factor.

\subsection{The geometrical factor of DGP}
The Einstein equations from Eq.~\ref{5Daction} and Eq.~\ref{4Daction}
are
\ba\label{Ein}
G_{(5)B}^{\,\,A}\equiv R_{(5)B}^{\,\,A}-\frac{1}{2}R g_{(5)B}^{\,\,A}
=\frac{8\pi}{M_{(5)}^3}T_{\,\,\,\,\,B}^{A}.
\ea
The tensor $T_{\,\,\,\,\,B}^{A}$ is composed of the energy momentum
tensors of the bulk and the brane and the scalar curvature of the induced
metric introduced in Eq.~\ref{4Daction}.
We assume that the bulk is empty and the brane has only presureless
matter components.
Then the total energy $\rho$ is
\ba\label{rho}
\rho=\rho_m-\frac{3M_{(4)}^2}{8\pi}\delta(y)\frac{\dot a^2}{a^2}.
\ea
The first term of right side of Eq.~\ref{rho} is
the matter density confined to the brane
and the second term comes from the scalar curvature of the brane 
\cite{deffayet00a}.

By considering the first integral of Eq.~\ref{Ein}
and the proper junction condition defining the derivative of $a(t,y)$
in terms of $y$ at crossing the surface $y=0$,
we get the following brane-FLRW equation 
\ba\label{hubble}
H^2+\epsilon \frac{H}{r_c}=\frac{8\pi}{3M_{(4)}^2}\rho_m,
\ea
where $\epsilon$ denotes signature ($+$, $-$). 
To be consistent with the observed cosmic acceleration, we choose the negative
sign for $\epsilon$. 
The negative sign implies
de Sitter-like brane embedding in the extra dimension.
The effect of the gravity leaking into
the extra dimension is significant at $H \sim 1/r_c$ 
but is negligible at $H \gg 1/r_c$ 
\cite{deffayet00a,deffayet01a}.

\begin{figure}[htbp]
  \begin{center}
    \plotone{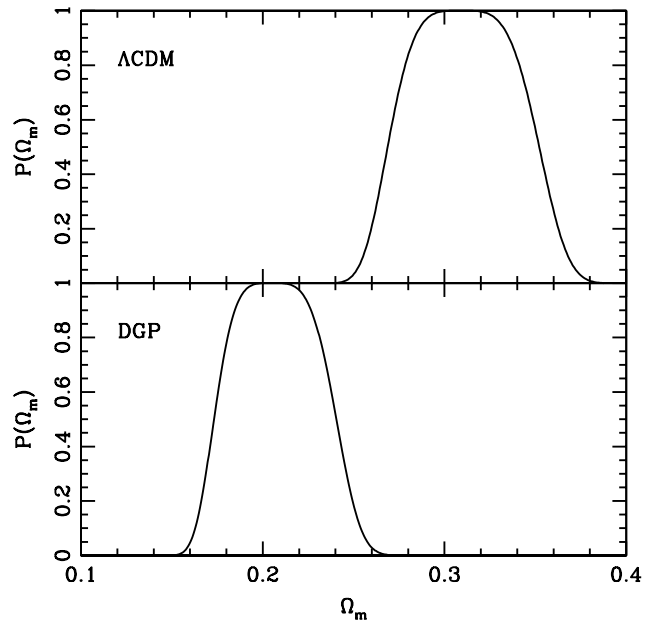}
    \caption{\label{f1} Probability distribution of $\Omega_m$ 
by fit the supernova data: 
top panel shows
the fit of $\Lambda$CDM and the bottom panel shows the fit
of DGP.
}
\end{center}
\end{figure}

From Eq.~\ref{hubble} we get 
\ba
H(a)=H_0\left(\sqrt{\Omega_c}+\sqrt{\Omega_c + \frac{\Omega_m}{a^3}}\right),
\ea
where the constant $\Omega_c$ is $\Omega_c=(1/2r_cH_0)^2$.
The flatness constraint is imposed to relate $\Omega_c$
to $\Omega_m$ as $\Omega_c=[(1-\Omega_m)/2]^2$.
Accordingly the crossover length scale is always greater than
the present Hubble horizon to avoid the negative value of $\Omega_m$
or the empty universe ($r_c \ga H^{-1}$).
There have already been attempts to fit the supernova data with DGP
\cite{avelino01,deffayet02b,dvali03a}.
We use the `golden set' of supernova measurements \cite{riess04}
and compare the likelihood function of $\Lambda$CDM model with DGP.
There is only a slight difference in the minimum $\chi^2$ between
$\Lambda$CDM and DGP, $\chi^2_{\rm min}=178$ for $\Lambda$CDM
and $\chi^2_{\rm min}=179$ for DGP. 
As far as the supernova experiment is concerned, 
both models are almost equally probable.
$\Lambda$CDM model prefers high $\Omega_m$ around 0.3
and DGP prefers low $\Omega_m$ around 0.2 (Fig.~\ref{f1}).

\subsection{The growth factor of DGP}\label{sec:core}
We choose the Gaussian normal longitudinal (GNL) gauge 
to fix the degrees of gauge freedom in 5D gravity.
In GNL gauge,
the extra dimension is not perturbed and any transition of energy from
the extra dimension to time coordinate or spatial coordinates is not allowed.
Then we have the usual (3+1)-dimensional longitudinal gauge on the brane
which leads to the perturbations in the time and spatial coordinates as
\ba
g_{00}&=&-1-2\Psi(t,\vec x,y=0),\nn\\
g_{ij}&=&a^2(t,y=0)\left[1+2\Phi(t,\vec x,y=0)\right]\delta_{ij}.
\ea
The 00 component of perturbed Einstein equation is given by 
\cite{mukohyama00a,deffayet02a}
\ba\label{00}
\frac{2}{a^2}\left[k^2\Phi+
3\dot a\left(\dot\Phi-\dot a\Psi\right)\right]
=\frac{1}{6}\left(\frac{8\pi}{M^3_{(5)}}\right)^2\rho\delta\rho,
\ea
where $\rho$ is given by Eq.~\ref{rho} and $\delta\rho$ is given by
\ba\label{drho}
\delta\rho=\delta\rho_m-\frac{8\pi}{3M^2_{(4)}}
\frac{2}{a^2}\left[k^2\Phi+
3\dot a\left(\dot\Phi-\dot a\Psi\right)\right].
\ea
Neglecting the contribution of velocity perturbations,
we write the Poisson equation  
by using Eq.~\ref{rho}, Eq.~\ref{00} and Eq.~\ref{drho} as
\ba\label{poisson}
k^2\Phi= \frac{4\pi}{M^2_{(4)}} a^2\rho_m
\frac{1}{1-\frac{1}{2r_cH}}
\delta_m,
\ea
where $\delta_m$ denotes the density contrast $\delta\rho_m/\rho_m$.
We do not consider $\epsilon=1$ case which is ruled out 
by the current measure distance.

The presence of Weyl component \cite{deffayet02a} does not deform 
the Poisson equation Eq.~\ref{poisson}.
The contribution of Weyl component is linearly added to the matter component
with suppression factor of $1/2r_cH$ at early time.
It could be important at later time when $1/2r_cH$ is close to unity.
But in GN coordinate gauge, the trace of Weyl component vanishes and the fluid
behaves like the radiation component.
If Weyl fluid behaves like the radiation component then the
clustering of that component is supressed and eventually it gaurantees
the approximation of smallness of Weyl contribution 
against the matter clustering.

In the DGP model, matter does not flow into the extra dimension,
i.e. there is no momentum component along the extra dimension $y$.
The perturbation length scales of interest
in this paper range approximately from 
$10^{-4}H_0^{-1}$ to $10^{-3}H_0^{-1}$. 
In this range, the gravity of density fluctuations is not dissipated into
the extra dimension significantly since the physical scales of those modes
are much greater than $1/r_c$.
Despite the presence of the extra dimension
the set of coupled continuity equations of matter is not altered.
In those intermediate scales, the coupled equations are
\ba\label{coupled}
\frac{d\delta_m}{d\eta}&+&ikv_m = 0, \nn\\
\frac{dv_m}{d\eta}&+&aHv_m =ik\Phi,
\ea
where we differentiate the matter perturbation $\delta_m$
and the velocity perturbation $v_m$ in terms of the comoving time $\eta$.

Second order differential equation from Eq.~\ref{coupled}
can be written as
\ba\label{deltam}
\frac{d^2\delta}{d\eta^2}+aH\frac{d\delta}{d\eta}-k^2\Phi=0.
\ea
By using Eq.~\ref{hubble} and Eq.~\ref{poisson},
the curvature perturbation $\Phi$ is given by the background expansion
rate $H(z)$ as
\ba\label{ptoH}
k^2\Phi=\frac{3}{2}a^2H^2
\frac{1-\frac{1}{r_cH}}{1-\frac{1}{2r_cH}}
\delta_m.
\ea
The curvature perturbation of DE is suppressed by 
the ratio of matter component to the gross energy in the universe,
but the curvature perturbation of DGP is suppressed by
the weakened gravity.
The difference in causes of suppression leads to the distinct 
evolution of growth factors of DE and DGP.

\section{model independent parametrization}
In this section we introduce a new parametrization to measure directly
the geometrical factor and the growth factor by tomographic cosmic shear.
First, we briefly discuss cosmic shear.

The size and shape of galaxies is altered by gravitational lensing.
The effect of lensing is described by the convergence $\kappa$
measuring the magnification or demagnification and the shear
components, $\gamma_1$ and $\gamma_2$, quantifying the distortion of
shape \cite{kaiser92,bartelmann01}.
The shear components can be inferred from measurement of galaxy
ellipticities which are composed of an intrinsic ellipticity 
and a lensing-induced ellipticity.
In the absence of correlations between the intrinsic ellipticities,
the rms error in the measurement of each shear component is
\ba\label{rms}
\sigma(\gamma_1)=\sigma(\gamma_2)=\gamma_{\rm rms}/\sqrt{N_{\rm pix}},
\ea
where $\gamma_{\rm rms}$ is the rms intrinsic shear of the galaxies
and $N_{\rm pix}$ is the number of galaxies in the pixel.

We assume that photometrically-determined redshifts can be used to sort
the source galaxies into eight redshift bins with $\Delta z=0.4$ 
from $z=0.0$ to $z=3.2$ \cite{hu99b}.
Maps of the shear components can be decomposed into even parity $E$ mode
and odd parity $B$ modes.
The signal contribution to the covariance of shear $E$ modes is correlated
across bins, but otherwise diagonal.
The $E$ modes shear, $\gamma_E$, power spectra are 
\ba\label{shearshear}
C_{l,\,ij}^{\gamma_E\gamma_E}=
\frac{\pi^2l}{2}\int dr r W(\bar r_i,r)W(\bar r_j,r)
\Delta_{\Phi}^2(k_{\bot},r),
\ea
where $\bar r_i$ denotes the angular diameter distance
to the median redshift of bin $i$.
The geometrical factor appears in the window function $W$
and the growth factor appears in dimensionless curvature power 
spectrum $\Delta_{\Phi}^2$.
We show how to measure directly those quantities in the following 
subsections.

\subsection{Parametrization of the angular diameter distance}
The lensing-induced ellipticities
are weighted by the mean location of sources in
the window function $W(\bar r,r)$ given by
\ba\label{win}
W(\bar r_i,r)=\frac{\bar r_i-r}{\bar r_i r}\,\,\,
({\rm for}\,\,r<\bar r_i,\,\,0\,\,{\rm otherwise}).
\ea
We directly parametrize the geometrical factor $\bar r_i$ 
instead of determining $\bar r_i$ by the cosmological parameters
depending on the model choice between DGP and DE.

\subsection{Parametrization of the growth factor}
The dimensionless curvature power spectrum in Eq.~\ref{shearshear}
is
\ba
\Delta_{\Phi}^2(k,r(z))=\frac{2\pi^2}{k^3} A_0 \,g_{\Phi}(z)^2 T(k,z)^2,
\ea
where $A_0$ is the primordial amplitude, 
$g_{\Phi}(z)$ is growth function of $\Phi$
and $T(k,z)$ is the transfer function.
Here we are interested in the intermediate scales 
from $10^{-4}H^{-1}_0$ to $10^{-3}H^{-1}_0$
where the perturbations of the dark energy 
do not grow significantly to influence
on the matter perturbations in DE, and the gravity of perturbations
does not dissipate into the extra dimension in DGP.
In those scales, the transfer function with negligible time dependence
can be determined without the detailed knowledge of DE or DGP,
and the departure from the linearity of density perturbations 
is not significant.

We parametrize the growth factor by discretizing its evolution
in each z bin as
\ba\label{eq:Fi}
F_i\equiv A_0 \frac{g(\bar z_i)^2}{a^2},
\ea
where the growth factor $F_i$ of matter perturbations 
is averaged inside each bin $i$
and is normalized to $A_0$ in the matter dominated era.
The curvature growth factor $g_{\Phi}$ is related to the matter perturbation
growth factor $g$ by Eq.~\ref{poisson}.

\begin{table}
\begin{center}
\begin{tabular}{ccccccc}
Experiment & $l^{\rm T}_{\rm max}$& $l^{\rm E,B}_{\rm max}$ & $\nu$ (GHz) & $\theta_b$ & $\Delta_T$ & $\Delta_P$\\
\tableskip\hline\tableskip
Planck      &2000 &2500  &  100 & 9.2' & 5.5 & $\infty$ \\
              & &   &  143 & 7.1' & 6  & 11 \\
              & &   &  217 & 5.0' & 13 & 27 \\
\tableskip\hline\tableskip
CMBpol       &2000  &2500 &  217 & 3.0' & 1  & 1.4 \\
\tableskip\hline
\end{tabular}
\end{center}
\caption{The specifications of CMB experiments}
\label{CMBspec}
\end{table}

\section{Experiments and Results}
We sort out the cosmological parameters into three groups:
primordial parameters, intermediate parameters and low redshift parameters.
The primordial parameters come from a generic parametrization
of structureless initial conditions.
The intermediate parameters contribute to leave a signature on
CMB acoustic peaks at last scattering surface.
The low redshift parameters affect the density fluctuations
at later times.
The primordial parameters consist of the primordial scalar amplitude $A_0$,
the scalar spectral index $n_S$ 
and the scalar running of the spectral index $\alpha_S$.
The intermediate parameters consist of the matter component $\omega_m$,
the baryon component $\omega_b$, the helium fraction $y_{He}$ and the angular
extent of sound horizon at the last scattering surface $\theta_S$.
The low redshift parameters consist of the neutrino mass
$m_{\nu}$, the reionization redshift $z_{\rm re}$ 
and all other parameters related
to the cosmic acceleration.

We consider the combination of CMB experiment with weak lensing survey
in order to simultaneously constrain all cosmological parameters.
The CMB angular power spectrum leads to tight constraints on 
the primordial parameters and intermediate parameters.
The power of constraints on all cosmological parameters
with CMB alone is shown in \cite{kaplinghat03},
and the constraints on the cosmological parameters 
by the combination of CMB and weak lensing are extensively
studied in \cite{takada03,song03b}.
The primordial parameters and intermediate parameters are tightly
constrained with the future CMB experiments
(see specifications in TABLE \ref{CMBspec}). 

The primordial amplitude $A_0$ can be precisely determined by CMB alone 
\cite{kaplinghat03}.
The overall amplitude shift of CMB power spectra also results from
the scattering of photons by ionized gas 
in the intergalactic medium.
Though both cosmological parameters $A_0$ and $z_{\rm re}$
are degenerate in most scales,
the degeneracy between $A_0$ and $z_{\rm re}$ is broken
by the reionization bump of CMB polarization anisotropies in large scales.
But the precision is limited by the unknown reionization history
\cite{kaplinghat02}.
The primordial scalar amplitude $A_0$ can be determined up to
1.7 percent level of accuracy in the absence of lensing contribution.

\begin{table}
\begin{center}
\begin{tabular}{cccc}
Experiment & ${\rm f_{sky}}$ &$\bar n_{\rm tot}$
& $\bar n_{\rm tot}/\gamma^2_{\rm rms}$\\ 
\tableskip\hline\tableskip
G2$\pi$    & 0.5  & 65 &  1900 \\
\tableskip\hline\tableskip
S3000      & 0.072 &  100 & 2920 \\
\tableskip\hline
\end{tabular}
\end{center}
\caption
{Weak lensing experimental parameters assumed.
Units for the total source sky density $\bar n_{\rm tot}$
is $1/{\rm arcmin}^2$ and the per-component rms shear ${\rm rms}$
is evaluated at $z=1$.} 
\label{shearspec}
\end{table}

For the future weak lensing surveys, we assume uniform coverage
which gives a diagonal noise covariance matrix ${\bf N}$,
\ba
N_{lm\,i,\,\,l'm'\,j}
=\frac{\gamma_{\rm rms}^2}{n_i}\delta_{ll'}\delta_{mm'}\delta_{ij}
\ea
where $n_i$ is the number of source galaxies per pixel in the given bin $i$. 
We use a ground-based survey G2$\pi$,
and a space-based surveys S3000.
The specifications for those surveys are shown in TABLE~\ref{shearspec}.
The advantages of space are not fully exploited in this analysis 
which is only using information on large-scale. 
The galaxy number distribution for the ground-based survey G2$\pi$ 
are inferred from observations with the Subaru telescope
\cite{nagashima02}. The analytic expression for this distribution
is well-matched with
\ba
dn/dz &\sim& z^{1.3}\exp\left[-(z/1.2)^{1.2}\right]\,\,\,\,\,(z<1)\nn\\
dn/dz &\sim& z^{1.1}\exp\left[-(z/1.2)^{1.2}\right]\,\,\,\,\,(z>1)
\ea
We normalize the total number of galaxies $\bar n_{\rm tot}$
as 65 per sq.arcmin after
reducing $dn/dz$ a half in the $1.2<z<2.5$ due to
inaccurate photometric redshift observation \cite{tyson02}.
For the space-based survey, the distribution function is
\ba
dn/dz &\sim& z^{2}\exp(-z/1.5)
\ea
with reaching higher limiting magnitudes and
having relatively accurate photometric redshifts \cite{massey03}.

The cosmological parameters considered in this paper are
$(A_0,n_S,\alpha_S,\omega_m,\omega_b,y_{He},\theta_S,z_{\rm re},m_{\nu})$
and ($\bar r_1$, ... ,$\bar r_8$,$F_1$, ... , $F_8$).
The Fisher matrix analysis is used to estimate the constraints on
all 25 cosmological parameters simultaneously.
We show the constraints on $\bar r_i$ and $F_i$ in the following subsections.

\subsection{Constraints on the angular diameter distance}
The extensive window function defined by $W(\bar r,r)\times T(k)$
is bell-shaped.
As $r$ is close to $\bar r_i$, 
the extensive window function decreases due to
the numerator term of $W$ in Eq.~\ref{win}.
And as $r$ is close to zero, 
the extensive window function also decreases since
the transfer function decreases at high $k=l/r$.
The variation of $\bar r_i$ contributes to shift
the peak point of the extensive window function.
Larger median distance leads to more contribution
of the smaller $k$ modes to cosmic shear correlation functions.

\begin{figure}[htbp]
  \begin{center}
    \plotone{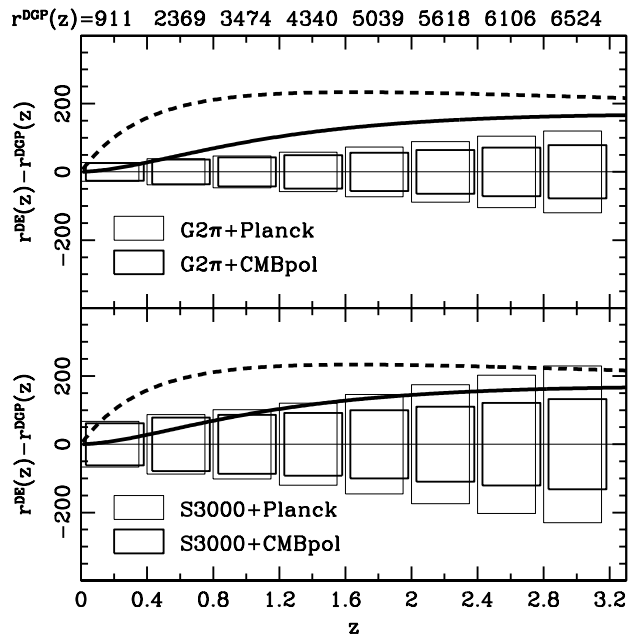}
    \caption{\label{f2} 
Constraints on the direct measurement of the angular diameter distance:
The error boxes are the level of accuracy to determine $\bar r_i$ in
each bin.
The solid curves represent the $\bar r_i$ of {\it Model DEr}
and the dash curves represent the $\bar r_i$ of {\it Model DEF}. 
}
\end{center}
\end{figure}

As is shown in Fig.~\ref{f2}, the angular diameter distance evolution
is well-reconstructed by tomographic cosmic shear.
The mean angular diameter distances $\bar r_i$
are directly measured in high accuracy around a few percentage level.

\subsection{Constraints on the growth factor}
The variation of $F_i$ has no impact on
the shear correlations in bin $j$ less than $i$.
The response to the variations of $F_i$ in bin $j=1$
is a unique signature of a parameter $F_1$.
Such a ladder structure helps to break degeneracy
among the eight $F_i$ parameters.
The constraints on $F_i$ at higher redshift bin
are weakened since fewer shear correlation power spectra
are influenced by $F_i$.


\begin{figure}[htbp]
\begin{center}
  \plotone{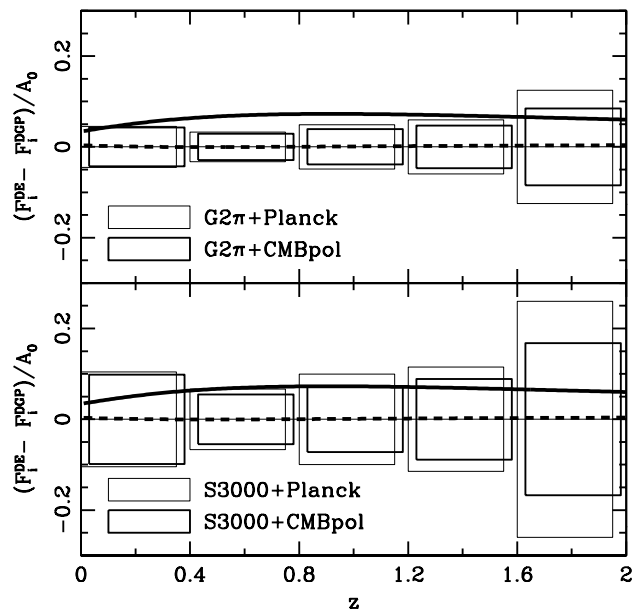}
  \caption{\label{f3} 
Constraints on the direct measurement of the growth factor:
The error boxes are the level of accuracy to determine $F_i$ 
as defined in Eq.~\ref{eq:Fi}.
The solid curves represent the $F_i$ of {\it Model DEr}
and the dash curves represent the $F_i$ of {\it Model DEF}.
}
\end{center}
\end{figure}

Based upon the measurement of the primordial scalar amplitude $A_0$
at the last scattering surface, $\sigma(\ln A_0)\sim 0.17$,
we proceed to construct the growth factor from higher redshift bins.
The growth factor history is precisely reconstructed
up to $z\sim 2$ with the ground-based surveys and up to $z\sim 1.2$ 
with the space-based surveys as shown in Fig.~\ref{f3}.

\section{Can we discriminate DGP from DE?}
The direct measurements of the eight $\bar r_i$ parameters
and the eight $F_i$ parameters allow us to
find the parameters that fit $\bar r_i$ and $F_i$.
In the DGP model, the $\bar r_i$ and $F_i$ are completely 
fixed by 9 other parameters 
$(A_0,n_S,\alpha_S,\omega_m,\omega_b,y_{He},\theta_S,z_{\rm re},m_{\nu})$
and the flatness constraint.
We assume that the fixed $\bar r_i$ and $F_i$ of DGP are fiducial values.
In the DE model, the $\bar r_i$ and $F_i$ are still variable by
the equation of state of dark energy $w(a)$ 
even with all constraints given by 9 other parameters and the flatness.
We parametrize $w(a)$ as $w+w_a(1-a)$ \cite{linder02}
and vary $(w,w_a)$ of DE to fit the fiducial $\bar r_i$ and $F_i$.
{\it Model DEr} denotes the best DE model to fit the fiducial $\bar r_i$ 
and {\it Model DEF} denotes the best DE model to fit the fiducial $F_i$.
In Fig.~\ref{f2} and Fig.~\ref{f3},
{\it Model DEr} and {\it Model DEF} provide
the solid curves and the dash curves respectively.

\begin{figure}[htbp]
  \begin{center}
    \plotone{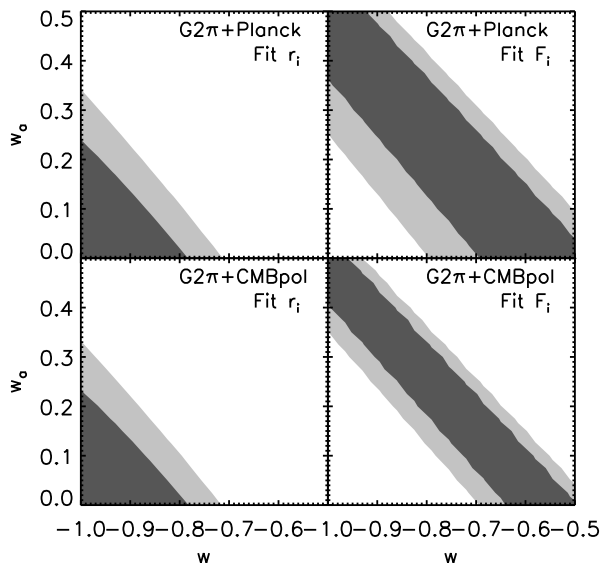}
    \caption{\label{f4} 
Contour plots showing constraints on the DE parameters with G2$\pi$,
assuming the true model DGP.
The contours in left panels cover DE parameter space
to fit $\bar r_i$ and
the contours in right panels cover DE parameter space
to fit $F_i$.
The fiducial model is the DGP model and we vary DE parameters
$(w,w_a)$ to fit the $\bar r_i$ and $F_i$.
The inner curves represent 1-$\sigma$ confidence level and 
the outer curves represent 2-$\sigma$ confidence level. 
}
\end{center}
\end{figure}

While the $F_i$ of {\it Model DEF} is nearly identical 
with the fiducial $F_i$, the $\bar r_i$ of {\it Model DEr} 
is distinct from the fiducial $\bar r_i$.
The angular diameter distance to the last scattering surface
at $z\sim1100$ is almost fixed by the prior information of 
$\theta_S$ and $\omega_m$. 
When we fix the angular diameter distance at $z\sim1100$,
we do not find any DE model which generates $\bar r_i$ nearly identical 
with the fiducial $\bar r_i$ by varying $(w,w_a)$.
By fitting $\bar r_i$,
DGP is able to be discriminated from DE models parametrized with $(w,w_a)$. 
However, any one of DE and DGP models is not excluded more than 
90\% confidence level with fitting $\bar r_i$ alone.
Also considering the generic aspect of DE, we can differently parametrize 
$w(a)$ to fit better the fiducial $\bar r_i$.
Thus it is not clear yet to discriminate DGP from DE
by fitting $\bar r_i$ alone
unless DGP is excluded by its poor fit to $\bar r_i$.

\begin{figure}[htbp]
  \begin{center}
    \plotone{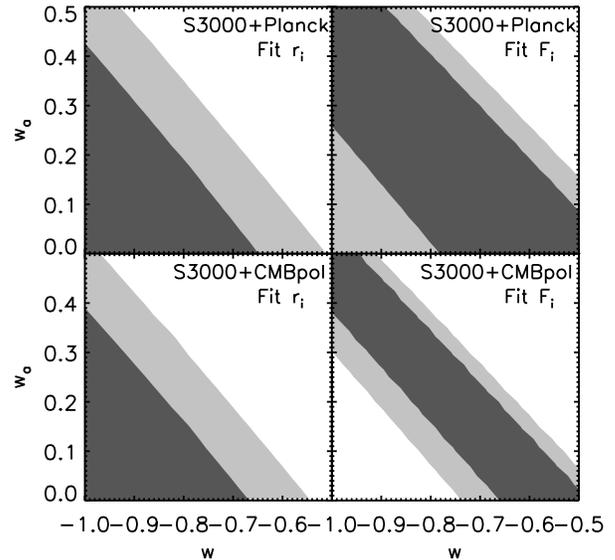}
    \caption{\label{f5} 
Contour plots showing constraints on the DE parameters with S3000,
assuming the true model DGP.
The contours in left panels cover DE parameter space
to fit $\bar r_i$ and
the contours in right panels cover DE parameter space
to fit $F_i$.
The fiducial model is the DGP model and we vary DE parameters
$(w,w_a)$ to fit the $\bar r_i$ and $F_i$.
The inner curves represent 1-$\sigma$ confidence level and 
the outer curves represent 2-$\sigma$ confidence level. 
}
\end{center}
\end{figure}

Next we consider fitting the $\bar r_i$ and $F_i$ simultaneously.
Despite a good fit with the fiducial $F_i$,
{\it Model DEF} is excluded by a poor fit with the fiducial $\bar r_i$.
The distinct suppression in the DGP growth factor leads to the separation
in two regions of the DE parameter space to fit the $\bar r_i$ and the $F_i$.
In Fig.~\ref{f4} and Fig.~\ref{f5}, 
we show the disagreement between two DE parameter spaces
in detail.
The ground-based cosmic shear survey G2$\pi$ clearly discriminate DGP
from DE around 95\% confidence level (Fig.~\ref{f4}).
And the space-based cosmic shear survey S3000 is able to see 
the difference between DE and DGP around 68\% confidence level (Fig.~\ref{f5}).
Although the space-based surveys in our analysis do somewhat worse
because of the smaller sky coverage, we must remind the reader that we
are focusing on science that can be done with information on the
larger angular scales.  Space-based observations will have
advantages with respect to ground--based observations at smaller
angular scales.

In conclusion, DGP is distinguishable from DE
by a cosmic shear survey.
The DGP model is not yet clearly ruled out by any experiment.
For instance, 
it is too early for us to rule out DGP by using $\sigma_8$ at this moment.
As shown in Fig.~\ref{f3}, the normalization influenced 
by DGP could be around $5\%$ less.
However
considering the current constraints on the primordial amplitude $A_0$
(more than $5\%$ uncertainty),
we are not able to discriminate DGP from DE by $\sigma_8$ alone
with current datasets.
We still have a chance to detect the extra dimension in the future.

The DGP growth factor predicted by \cite{lue04} excluded the DGP model by more
suppression than ours.
Lue et. al. \cite{lue04} derived the gravitational perturbations
of spherically symmetric clustered mater sources on the cosmological
background.
However the current solution is possibly missing the significant contribution
of quadrature contribution (private communication with Roman Scoccimarro)
and we are not able to compare both different methods at this moment.

In case the future cosmic shear surveys prefer DGP to DE,
then we will also be able to precisely determine the crossover length scale
$r_c$.
The knowledge of $r_c$ will lead to the discovery of 
a fundamental energy scale
and mark the beginning of experimental exploration of extra dimensions.

\vspace{.3cm}
\noindent {\it Acknowledgments} :
This work is supervised by Lloyd Knox as a partial fulfillment
of PhD dissertation. We appreciate helpful comments from
Nemanja Kaloper, Manoj Kaplinghat, 
Jin-Young Kim, Arthur Lue, Roman Scoccimarro and Lorenzo Sorbo.
This material is based upon work supported by the National Science
Foundation under Grant No. 0307961
and U.S. Dept. of Energy contract DE-FG02-90ER-40560.

\bibliography{dgpbib}

\end{document}